\documentclass[
12pt,
superscriptaddress,
showpacs, showkeys,
amsmath,amssymb,
aps,
pre,
floatfix,
]{revtex4-1}

\usepackage{hyperref}
\usepackage{graphicx}
\usepackage{amsmath}
\usepackage{amsfonts}
\usepackage{amssymb}
\usepackage{yfonts}
\usepackage{bm}
\usepackage{siunitx}
\usepackage{enumitem}
\usepackage{xcolor}

\begin{document}

\title{Nature of Intrinsic Uncertainties in Equilibrium Molecular Dynamics Estimation of Shear Viscosity\\for Simple and Complex Fluids}

\author{Kang-Sahn Kim}
\affiliation{Department of Chemistry, Korea Advanced Institute of Science and Technology (KAIST), \\ Daejeon 34141, Republic of Korea}

\author{Myung Hoon Han}
\affiliation{Department of Chemistry, Korea Advanced Institute of Science and Technology (KAIST), \\ Daejeon 34141, Republic of Korea}

\author{Changho Kim}
\email{ckim103@ucmerced.edu}
\affiliation{Computational Research Division, Lawrence Berkeley National Laboratory, \\ Berkeley, California 94720, USA}
\affiliation{Applied Mathematics, University of California, \\ Merced, California 95343, USA}

\author{Zhen Li}
\affiliation{Division of Applied Mathematics, Brown University, \\ Providence, Rhode Island 02912, USA}

\author{George Em Karniadakis}
\affiliation{Division of Applied Mathematics, Brown University, \\ Providence, Rhode Island 02912, USA}

\author{Eok Kyun Lee}
\email{eklee@kaist.ac.kr}
\affiliation{Department of Chemistry, Korea Advanced Institute of Science and Technology (KAIST), \\ Daejeon 34141, Republic of Korea}

\date{\today}

\begin{abstract}
We study two types of intrinsic uncertainties, statistical errors and system size effects, in estimating shear viscosity via equilibrium molecular dynamics simulations and compare them with the corresponding uncertainties in evaluating the self-diffusion coefficient.
Uncertainty quantification formulas for the statistical errors in the shear-stress autocorrelation function and shear viscosity are obtained under the assumption that shear stress follows a Gaussian process.
Analyses of simulation results for simple and complex fluids reveal that the Gaussianity is more pronounced in the shear-stress process (related to shear viscosity estimation) compared with the velocity process of an individual molecule (related to self-diffusion coefficient).
At relatively high densities corresponding to a liquid state, we observe that the shear viscosity exhibits complex size-dependent behavior unless the system is larger than a certain length scale, beyond which reliable shear viscosity values are obtained without any noticeable scaling behavior with respect to the system size.  
We verify that this size-dependent behavior is configurational and relate the characteristic length scale to the shear-stress correlation length. 
\end{abstract}

\maketitle{}

\section{Introduction}


Shear viscosity is an essential transport property, which measures internal resistance of fluid.
Its theoretical estimation dates back to the early days of molecular dynamics (MD) simulations in 1970s~\cite{AlderGass1970, AlderWainwright1970, WainwrightAlder1971, LevesqueVerletKurkijarvi1973}.
Thereafter, numerous MD simulation studies have been performed for various fluid systems ranging from simple fluids~\cite{SchoenHoheisel1985, LevesqueVerlet1987, Erpenbeck1988,MeierLaesecke2004, Petravic2004b, FenzMryglod2009, IsobeAlder2012}, water~\cite{YehHummer2004, FengWong2009, GonzalezAbascal2010, SongDai2010, FanourgakisMedina2012, TaziBotan2012}, ionic liquids~\cite{BhargavaBalasubramanian2005, Rey-CastroVega2006, PicalekKolafa2009, IshiiKasai2015, NieszporekNieszporek2016}, polymer melts~\cite{PuscasuTodd2010, GuoChung2008, LeTodd2009, Son2009}, liquid metals~\cite{KressCohen2011, LevashovMorris2011, LevashovMorris2013, MeyerXu2016}, and blood~\cite{FedosovPan2011}.
Accurate and precise MD estimation of shear viscosity is important both from the practical perspective, e.g.~\cite{AlfeGillan1998}, and theoretical perspective, e.g.~\cite{IsobeAlder2009,IsobeAlder2010,ChoiHanKimTalknerKideraLee2017}.
Moreover, its importance has been recently recognized in the development of coarse-grained models~\cite{LiBianYangKarniadakis2016} and force fields~\cite{MorawietzSingraberDellagoBehler2016}, where a new fluid model is assessed by how closely it reproduces correct dynamic fluid properties.


In equilibrium MD simulations, transport coefficients are computed using the Green--Kubo formulas~\cite{Green1954,Kubo1957}.
The shear viscosity coefficient is expressed as $\eta=\lim_{t\rightarrow\infty}\eta(t)$ with the time-dependent shear viscosity
\begin{equation}
\label{eq:gk_eta_t}
    \eta(t) = \frac{V}{k_\mathrm{B} T} \int_{0}^t \langle p_{xy}(0) p_{xy}(t') \rangle dt' .
\end{equation}
Here, $\langle p_{xy}(0)p_{xy}(t)\rangle$ is the shear-stress autocorrelation function (SACF), where $p_{xy}$ denotes the $xy$-component of the stress tensor, and the brackets indicate an equilibrium average.
$V$ and $T$ indicate the volume and temperature of the system and $k_{\mathrm{B}}$ denotes Boltzmann's constant, respectively.
Alternatively, the shear viscosity can also be calculated from the long-time slope of the mean squared difference of the Helfand moment through the generalized Einstein relation~\cite{Helfand1960,ViscardyServantie2007}.
In nonequilibrium MD simulations, an external perturbation is applied to fluid systems and the shear viscosity is estimated from the resulting steady states~\cite{Erpenbeck1984,Heyes1986,EvansMorriss1989,FerrarioCiccotti1991,Muller-Plathe1999,BackerLowe2005} or time-transient behavior~\cite{CiccottiJacucci1979,AryaMaginn2000}.
In addition, approaches based on Onsager's thermodynamic formalism~\cite{PalmerSpeck2017} and the large deviation theory~\cite{GaoLimmer2017} have been recently proposed.


\begin{figure}
    \includegraphics[width=\textwidth]{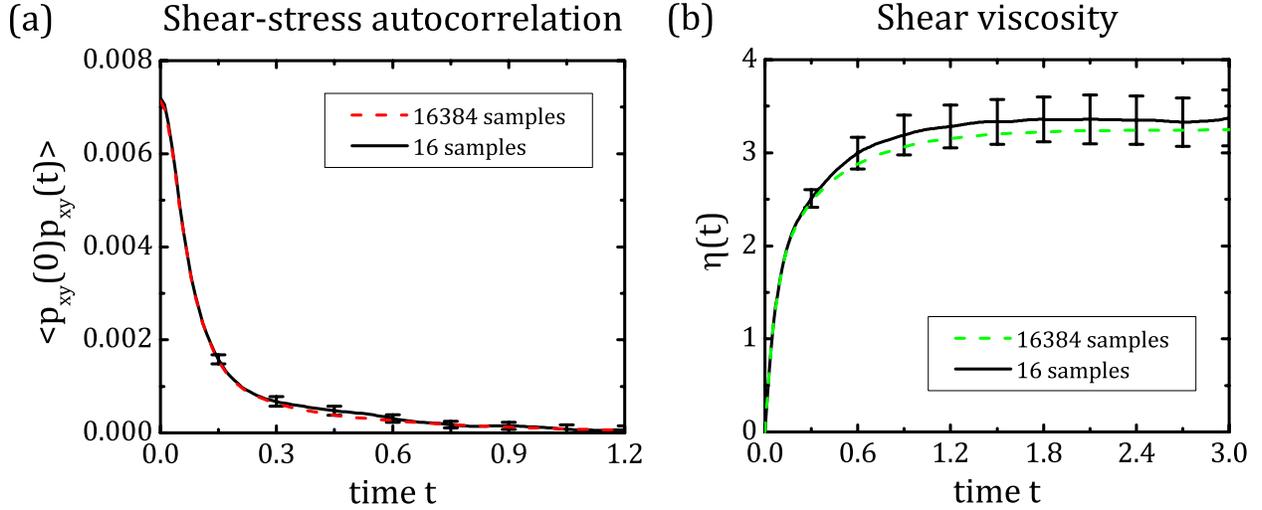}
    \caption{(a) SACF, $\langle p_{xy}(0)p_{xy}(t)\rangle$, and (b) time-dependent shear viscosity, $\eta(t)$, of an LJ fluid.
    Results obtained from 16 samples (depicted by black solid lines) and 16384 samples (dashed colored lines) are compared.
    For the 16-sample results, error bars corresponding to two standard deviations are drawn.
    The error bars of the 16384-sample results are not drawn since they are too small to be visible.
    Simulation details are provided in Section~\ref{sec_num_method}.}
\label{fig:scf_vis-LJ}
\end{figure}

\begin{figure}
    \includegraphics[width=\textwidth]{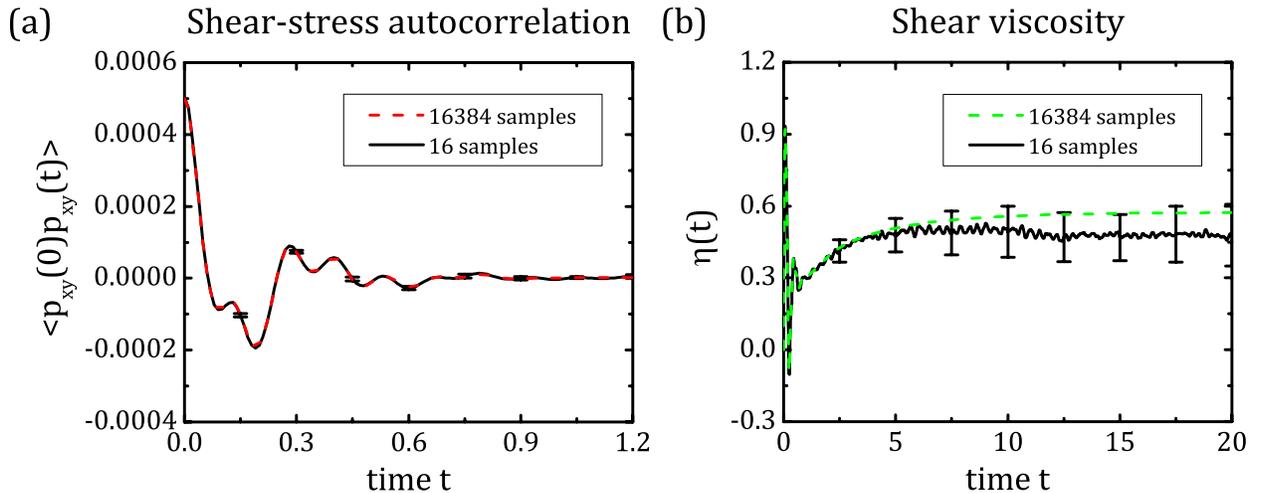}
    \caption{(a) SACF and (b) time-dependent shear viscosity of a star-polymer melt.
    Results obtained from 16 samples are depicted by black solid lines with error bars corresponding to two standard deviations, whereas results from 16384 samples are depicted by dashed colored lines.
    For the simulation details, see Section~\ref{sec_num_method}.}
\label{fig:scf_vis-starpoly}
\end{figure}


In the Green--Kubo method, two main sources of uncertainty hinder the accurate estimation of transport coefficients. 
The first originates from insufficient sampling.
While any quantity obtained from MD simulations is blurred by statistical errors, a characteristic feature in the Green--Kubo method is that statistical errors accumulate through time integration, making it difficult to verify the convergence of the Green--Kubo integral~\cite{JonesMandadapu2012,KimBorodinKarniadakis2015}.
Figures~\ref{fig:scf_vis-LJ} and \ref{fig:scf_vis-starpoly} demonstrate these difficulties when the shear viscosity is estimated from insufficient sampling averages for a Lennard-Jones (LJ) fluid and a star-polymer melt, respectively. 
Even if the SACF appears to show barely meaningful values beyond a certain time due to statistical errors, its time integral $\eta(t)$ needs to be computed up to time $t^*$ such that $\eta(t^*)\approx\eta$. 
However, $t^*$ is not known a priori and is hard to determine from the data with insufficient sampling.
Figure~\ref{fig:scf_vis-starpoly} clearly shows that a good result from the SACF may not be sufficient to estimate $t^*$.
Thus, the resulting value of $\eta$ may be inaccurate, especially for a complex fluid in which constituent molecules have various internal motions.

The statistical uncertainty is usually quantified in terms of the standard error (i.e. the standard deviation of the statistical error).
For the number of independent sample trajectories $\mathcal{N}$ and the length of each sample trajectory $\mathcal{T}$, Zwanzig and Ailawadi~\cite{ZwanzigAilawadi1969} and Frenkel~\cite{Frenkel1980} have shown that the standard error in an averaged quantity decreases proportional to $\mathcal{N}^{-1/2}$ and $\mathcal{T}^{-1/2}$.
Jones and Mandadapu~\cite{JonesMandadapu2012} have extended these analyses to obtain an upper bound of the standard error in the time integral of an autocorrelation function.
Recently, some of the co-authors of the present paper have derived general formulas estimating the standard errors in an autocorrelation function and its time integral~\cite{KimBorodinKarniadakis2015}.
These analyses are all based on the assumption that the underlying process is Gaussian.
Thus, any higher moment can be calculated from its first and second moments.
While the Gaussian process approximation (GPA) is believed to hold well for the description of various physical and chemical stochastic processes~\cite{Fox1978}, its validity needs to be confirmed for each process.
In Ref.~\cite{KimBorodinKarniadakis2015}, through an extensive and systematic MD study, the validity of the GPA was tested for the velocity process of a tagged particle in a three-dimensional simple fluid.
Consequently, the statistical error estimates for the velocity autocorrelation function (VACF) and the self-diffusion coefficient were validated.


Finite system size effect, resulting from the use of artificial boundaries such as periodic boundary conditions to mimic an infinite system, is another source of uncertainty.
Among transport coefficients, the self-diffusion coefficient of a three-dimensional fluid is the one for which the system size effects have been most thoroughly investigated.
Theoretical investigations based on the hydrodynamic theory have revealed that, for a simulation box with side length $L$, the leading-order correction term in the self-diffusion coefficient has an $L^{-1}$ scaling behavior~\cite{Hasimoto1959, DunwegKremer1993, YehHummer2004}.
Subsequently, this scaling behavior has been verified for various fluid systems, including LJ fluids~\cite{YehHummer2004, OhtoriIshii2015a, OhtoriIshii2015b}, Weeks--Chandler--Andersen (WCA) fluids~\cite{DunwegKremer1993, Heyes2007, Nasrabad2008}, hard-sphere fluids~\cite{Heyes2007, HeyesCass2007}, water~\cite{YehHummer2004, TaziBotan2012}, ionic liquids~\cite{GablSchroder2012, NieszporekNieszporek2016}, polymer melts~\cite{DunwegKremer1993,KuhneKrack2009}, and molten alkali halides~\cite{IshiiKasai2015}.
For the self-diffusion coefficient of a two-dimensional fluid, known to diverge logarithmically with increasing $L$~\cite{KeyesLadanyi1975, DonevBellFuenteGarcia2011, ChoiHanKimTalknerKideraLee2017}, system size effects have been recently analyzed based on the long-time tail of the VACF~\cite{ChoiHanKimTalknerKideraLee2017}, for which a valid hydrodynamic description is available~\cite{HanKimTalknerKarniadakisLee2018}.

Shear viscosity is known to be less dependent on system size than self-diffusion coefficient in three-dimensional fluid systems.
However, the existence of any clear scaling behavior is still debatable since most observations were based on limited computational results with considerable statistical uncertainties.
While size effects have been reported insignificant (within statistical uncertainties) for LJ fluids~\cite{DunwegKremer1993, HolianEvans1983, YehHummer2004}, water~\cite{YehHummer2004}, liquid sodium~\cite{MeyerXu2016}, liquid iron~\cite{LevashovMorris2011}, and molten alkali halides~\cite{IshiiKasai2015}, a weak scaling law, proportional to $L^{-3}$, has been presumed for the correction term to estimate the shear viscosity of the infinite system~\cite{Erpenbeck1988, MeierLaesecke2004, Heyes2007}, yet without theoretical substantiation.

Different system size effects on self-diffusion coefficient and shear viscosity suggest that the latter should have an additional mechanism causing the size effects other than the one with the hydrodynamic origin~\cite{ErnstHauge1971}.
The shear-stress relaxation due to configurational changes is such a mechanism.
In fact, the slow structural relaxation in dense fluids leads to ``molasses tail'' (stretched exponential decay)~\cite{IsobeAlder2009, IsobeAlder2010, IsobeAlder2012}, which is different from the long-time tail decaying algebraically~\cite{ErnstHauge1971}.
By defining a shear-stress correlation length measuring the size of a region in a liquid that can rearrange independent of environment, Petravic~\cite{Petravic2004a, Petravic2004b} has demonstrated that complex (i.e., non-scaling) size effects observed in small dense liquid systems can be understood through a concept similar to ``cooperatively rearranging regions'' of the Adam--Gibbs theory of glass transition.
B\"uchner and Heuer~\cite{BuchnerHeuer1999} have associated system size effects with the limited number of inherent structures (i.e., local energy minima) in finite systems.


The present paper investigates the nature of the two aforementioned types of uncertainties inherent in the estimation of shear viscosity.
To this end, we perform an extensive and systematic MD study for LJ fluid and star-polymer melt models and compare the case of the shear viscosity estimation with that of the self-diffusion coefficient estimation.
For the statistical uncertainty, we validate the uncertainty quantification formulas and investigate the origin of pronounced Gaussianity observed in the shear viscosity estimation.
For the uncertainty due to finite system size, we observe the system size-dependent behavior of shear viscosities of the two models and investigate their relations to atomic and molecular rearrangements.
We also propose an entity from which Petravic's shear-stress correlation length~\cite{Petravic2004a, Petravic2004b} can be readily estimated.
We impose high-density and ambient-temperature conditions so that the fluid models are in a typical liquid state.
The knowledge acquired through this study is practically useful for the accurate estimation of shear viscosity and theoretically important for the understanding of the underlying shear-stress relaxation process in a fluid.
In a broader context, it contributes to the recent advancement of uncertainty quantification techniques for MD simulation studies~\cite{AngelikopoulosPapadimitriou2012, RizziNajmDebusschereEtAl2012a, RizziNajmDebusschereEtAl2012b, RizziJonesDebusschereKnio2013a, RizziJonesDebusschereKnio2013b, PatroneDienstfreyBrowningEtAl2016, LeiYangLiKarniadakis2017} by providing more accurate characterization of intrinsic uncertainties in MD simulations~\cite{MaitreKnio2015, PatroneDienstfreyBrowningEtAl2016}.


The rest of the paper is organized as follows.
In Section~\ref{sec_background}, we briefly review statistical uncertainty quantification formulas and theoretical accounts of the system size effects in the estimation of transport coefficients.
In Section~\ref{sec_num_method}, we present the details of our MD simulations for the LJ fluid and star-polymer melt models. 
In Section~\ref{sec_statisticalErrors}, we present the statistical error analysis regarding the evaluation of shear viscosity.
In Section~\ref{sec_finiteSystemSizeEffect}, we provide the analysis on the system size-dependent behavior of shear viscosity. 
We conclude the paper by providing a summary and discussion in Section~\ref{sec_conclusion}.

\section{\label{sec_background}Background}

\subsection{\label{subsec_stat_uq_exp}Statistical Uncertainty Quantification}

Here we summarize the statistical uncertainty formulas~\cite{KimBorodinKarniadakis2015} for the Green--Kubo method.
According to the Green--Kubo relation, the time-dependent transport coefficient, $\gamma (t)$, is expressed as a time integral of the autocorrelation function of a corresponding process, $a(t)$,
\begin{equation}
\label{eq:gkRelation}
    \gamma(t) = A \int_{0}^{t} \langle a(0) a(t') \rangle dt' ,
\end{equation}
where $A$ denotes a prefactor and $\gamma = \lim_{t\rightarrow\infty} \gamma(t)$.
In practice, the time autocorrelation function, $C(t)=\langle a(0)a(t) \rangle$, is estimated from sample trajectories through the \textit{ensemble average} of \textit{time average}, which can be expressed as
\begin{equation}
\label{eq:est-corrFunc}
    \hat{y} (t) = \frac{1}{\mathcal{N}} \sum_{k=1}^{\mathcal{N}} \frac{1}{\mathcal{T}} \int_{0}^{\mathcal{T}} a^{(k)} (t') a^{(k)} (t' + t) dt' .
\end{equation}
Here, $\mathcal{N}$ and $\mathcal{T}$ denote the number of independent sample trajectories and the length of each sample trajectory, respectively.
The superscript $^{(k)}$ indicates that the quantity is obtained from the $k$th sample trajectory.
Accordingly, the estimator, $\hat{z}(t)$, of $\gamma(t)$ is expressed as
\begin{equation}
\label{eq:est-transCoef}
    \hat{z} (t) = A \int_{0}^{t} \hat{y} (t') dt' .
\end{equation}

The quantities of interest are the standard errors, $\sigma_a (t)$ and $\sigma_\gamma(t)$, in the estimators, $\hat{y}(t)$ and $\hat{z}(t)$, respectively.
In other words, for the statistical errors, $\varepsilon_a (t) = \hat{y} (t) - \langle \hat{y} (t) \rangle$ and $\varepsilon_\gamma (t) = \hat{z} (t)- \langle \hat{z} (t) \rangle$, we want to quantify their error levels by computing $\sigma_a^2 (t) = \langle \varepsilon_a^2 (t) \rangle$ and $\sigma_\gamma^2 (t) = \langle \varepsilon_\gamma^2 (t) \rangle$.
Since $\varepsilon_\gamma$ is correlated with $\varepsilon_a$ via Eq.~\eqref{eq:est-transCoef}, $\sigma_\gamma (t)$ is expressed in terms of the error correlation function, $\langle \varepsilon_a (t') \varepsilon_a (t'') \rangle$, of $\hat{y}$:
\begin{equation}
\label{eq:errorCorr}
    \sigma_\gamma^2 (t) = \int_0^t dt' \int_0^t dt'' \langle \varepsilon_a (t') \varepsilon_a (t'') \rangle .
\end{equation}
While Eq.~\eqref{eq:errorCorr} is exact, we note that the error correlation function is a fourth-order correlation function of $a(t)$, and its direct calculation through MD simulation is computationally impractical.

Under the GPA, i.e., $a(t)$ is a Gaussian process~\footnote{
A process $X_t$ is Gaussian if and only if, for any finite set of time points $\{ t_1, \ldots{}, t_k \}$, the joint distribution of $\{ X_{t_1}, \ldots{}, X_{t_k} \}$ follows a multivariate normal distribution.
Note that $X_t$ may not be a Gaussian process even if the marginal distribution of $X_{t_i}$ is Gaussian at each time $t_i$.}, 
the error correlation function is approximated using the property that higher-order moments of a Gaussian process can be expressed in terms of the first two lowest-order moments.
The approximation
\begin{equation}
\label{eq:gpa}
    \langle a(0) a(t_1) a(t_2) a(t_3) \rangle \approx C(t_1) C(t_3 - t_2) + C(t_2) C(t_3 - t_1) + C(t_3) C(t_2 - t_1)
\end{equation}
yields the following expressions for the respective standard errors in $C(t)$ and $\gamma (t)$:
\begin{align}
\label{eq:err-corrFunc}
    &\sigma_a (t) = \sqrt{ \frac{1}{\mathcal{NT}} \int_{-\infty}^{\infty} \! d\tau \Big[ C^2 (\tau) + C(\tau - t) C(\tau + t) \Big] },\\
\label{eq:err-transCoef}
    &\sigma_\gamma (t) = A \sqrt{ \frac{1}{\mathcal{NT}} \int_{-\infty}^{\infty} \! d \tau \left[ C(\tau) \int_{0}^{t} \! dt' \int_{\tau}^{\tau + t} \! dt'' C(t' - t'') + \int_{\tau}^{\tau + t} \! C(t') dt' \int_{\tau - t}^{\tau} \! C(t') dt' \right] } .
\end{align}

In the case of shear viscosity, $a(t)$ and $A$ correspond to $p_{xy} (t)$ and $V (k_\mathrm{B} T)^{-1}$, respectively.
Hence, once the SACF is computed from MD simulations, both the standard errors in the SACF and shear viscosity can be estimated from Eqs.~\eqref{eq:err-corrFunc} and \eqref{eq:err-transCoef}.
For the self-diffusion coefficient defined as $D=\lim_{t\rightarrow\infty} D(t)$ with 
\begin{equation}
\label{eq:gk_diff_t}
    D(t)=\int_0^t \langle v_x(0)v_x(t')\rangle dt',
\end{equation}
$a(t)$ corresponds to the velocity component, $v_x(t)$, of a tagged particle with $A=1$.

The following observations can be made from Eqs.~\eqref{eq:err-corrFunc} and \eqref{eq:err-transCoef}.
First, the standard errors, $\sigma_\bullet$, are proportional to $(\mathcal{NT})^{-1/2}$.
Hence, one can define the \emph{normalized} standard errors
\begin{equation}
\label{normSE}
\widetilde{\sigma}_\bullet = \sqrt{\mathcal{NT}}\sigma_\bullet.
\end{equation}
Second, for a $C(t)$ decaying to zero, one can easily show that $\sigma_a(t)$ converges to a positive constant, satisfying
\begin{equation}
\label{eq:asymptACF}
\lim_{t\rightarrow\infty} \sigma_a(t) = \frac{1}{\sqrt{2}}\sigma_a(0).
\end{equation}
Third, as shown in the Appendix, the standard error of a time-dependent transport coefficient grows asymptotically as fast as $\sqrt{t}$, satisfying
\begin{equation}
\label{eq:asymptCoef}
\lim_{t\rightarrow\infty}\frac{\widetilde{\sigma}_\gamma(t)}{2\gamma\sqrt{t}}=1.
\end{equation}
An identical asymptotic expression has been derived in Appendix~A of Ref.~\cite{JonesMandadapu2012}.

\subsection{Transport Coefficients in Finite Periodic Systems}

\subsubsection{Self-diffusion Coefficient}

The self-diffusion coefficient, $D_L$, of a three-dimensional fluid in a periodic simulation box with side length $L$ is known to satisfy the following relation with the self-diffusion coefficient, $D_\infty$, in an infinite system~\cite{Hasimoto1959, DunwegKremer1993, YehHummer2004}:
\begin{equation}
\label{eq:diffCoef}
    D_L = D_\infty - \frac{2.837 k_{\mathrm{B}} T}{6 \pi \eta L} .
\end{equation}
A physical interpretation of this relation is as follows.
The diffusion of a tagged particle is influenced by the long-range hydrodynamic interaction between the tagged particle and the surrounding fluid~\cite{AlderWainwright1970, ChoiHanKimTalknerKideraLee2017, HanKimTalknerKarniadakisLee2018}.
In a finite periodic system, the hydrodynamic interaction developing in a given cell is interrupted by that developing in neighboring images, resulting in system size dependence.

\subsubsection{Shear Viscosity}

For a periodic molecular system where atoms interact pairwise via central pair-potentials, $\phi_{ij}(r)$, the stress tensor is expressed as~\cite{HolianEvans1983, ThompsonPlimptonMattson2009}
\begin{equation}
\label{eq:virial}
    p_{\alpha \beta} = \frac{1}{V} \left[ \sum_{i} m_i v_{i,\alpha} v_{i,\beta} - \frac12 \sum_i \sum_{j \ne i} \frac{r_{ij,\alpha}r_{ij,\beta}\phi_{ij}'(r_{ij})}{r_{ij}} \right],
\end{equation}
where $m_i$ and $v_{i,\alpha}$ are the mass and $\alpha$-component ($\alpha=x,y,z$) of the velocity vector of the $i$th atom, respectively.
$r_{ij}$ and $r_{ij,\alpha}$ are the interatomic distance and $\alpha$-component of the displacement vector from atom $j$ to atom $i$, respectively.
From Eq.~\eqref{eq:virial}, $p_{\alpha\beta}$ is decomposed into kinetic and potential contributions.
While the resulting kinetic contributions in the SACF and shear viscosity are believed to have a hydrodynamic origin as theoretically studied~\cite{ErnstHauge1971}, recent MD studies~\cite{IsobeAlder2009, IsobeAlder2010, IsobeAlder2012} have revealed that the potential counterpart due to structural relaxation becomes more significant in dense fluids. 

To demonstrate that a critical system size exists below which the shear viscosity estimation suffers from significant system size effects, Petravic~\cite{Petravic2004a, Petravic2004b} has shown that a small dense liquid system under \emph{shifted} periodic boundary conditions (i.e., subject to constant strain) can sustain \emph{unphysical} shear-stress.
For the total unrelaxed shear-stress, defined as
\begin{equation}
\label{eq:unrelaxedStress}
    \kappa = \left[ \sum_{\alpha} \sum_{\beta} \langle \Pi_{\alpha\beta} \rangle \langle \Pi_{\beta\alpha} \rangle \right]^{1/2}
\end{equation}
with $\Pi_{\alpha\beta}$ denoting the traceless stress tensor, an overall yet \emph{non-monotonic} decay of $\kappa$ was observed with increasing system size.
The shear-stress correlation length was defined as a characteristic side length of the periodic cell for which $\kappa$ vanishes irrespective of the boundary strain.
Correspondingly, this length scale was related to that of a liquid subsystem (or cooperatively rearranging region) over which the shear-stress fluctuations are spatially correlated.
The observed system size effect was shown to be configurational, which results from the scarcity of possible configurations in a small dense system.

\section{\label{sec_num_method}MD Simulations}

As a simple fluid model, we consider a three-dimensional LJ fluid at number density $\rho_\mathrm{LJ} = 0.8442$ and temperature $T_\mathrm{LJ} = 0.722$.
The interaction between particles is described via the LJ potential given by
\begin{equation}
V_{\mathrm{LJ}}(r) = 4\varepsilon \left[ {\left( \frac{\sigma}{r} \right) }^{12} - {\left( \frac{\sigma}{r} \right) }^{6} \right] .
\end{equation}
We use the reduced units of mass, length, and energy, i.e., $m=\sigma=\varepsilon=1$ with $k_\mathrm{B}=1$.
The cutoff radius of the potential is set to $r_\mathrm{c}=2.5$.
MD simulations of various system sizes were performed under periodic boundary conditions.
The smallest system has an $N_\mathrm{LJ} = 128$ particles and the largest with $N_\mathrm{LJ} = 65536$ particles.
The side length of a cubic simulation box is accordingly determined as $L=(N_\mathrm{LJ}/\rho_\mathrm{LJ})^{1/3}$. 
$NVE$ simulations were conducted using the velocity Verlet algorithm implemented in LAMMPS~\cite{Plimpton1995} with time step $\Delta t=0.002$.
Each equilibrium sample was obtained through equilibration for period $\mathcal{T}_\mathrm{equil} = 10^5 \Delta t = 200$.
The subsequent production run was performed for period $\mathcal{T}=10^5 \Delta t = 200$.  
For each set of simulation parameters, a total of $\mathcal{N}=16384$ sample trajectories were calculated.

We also consider a star-polymer melt model used in Ref.~\cite{LiBianCaswellKarniadakis2014} and adopt the same notations and parameters therein.
Each star-polymer has $N_\mathrm{a}$ arms with $N_\mathrm{b}$ beads per arm.
The arms are linked to a central bead.
Hence, there are $N_\mathrm{c}=N_\mathrm{a}N_\mathrm{b}+1$ beads per molecule.
We set $N_\mathrm{a}=10$ and vary $N_\mathrm{b}=1,2,3$ to investigate the influence of effective molecular size.
Typical configurations of star-polymers with $N_\mathrm{c}=11,21,31$ are illustrated in Fig.~1 of Ref.~\cite{LiBianCaswellKarniadakis2014}.
Excluded volume interactions between beads are described via the WCA potential given by
\begin{equation}
V_{\mathrm{WCA}}(r) = \left\{ \begin{array}{cc} 4\varepsilon \left[ {\left( \frac{\sigma}{r} \right) }^{12} - {\left( \frac{\sigma}{r} \right) }^{6} + \frac{1}{4} \right], & r \leq 2^{1/6} \sigma , \\ 0, & r > 2^{1/6} \sigma . \end{array} \right.
\end{equation}
We use the reduced units of mass, length, and energy, i.e., $m=\sigma=\varepsilon=1$ with $k_\mathrm{B}=1$.
Bond interactions are given by the finitely extensible nonlinear elastic (FENE) potential
\begin{equation}
V_{\mathrm{FENE}}(r) = \left\{ \begin{array}{cc} - \frac{1}{2} k r_{0}^{2} \phantom{1} \mathrm{ln} \left[ 1 - {( \frac{r}{r_0} )}^{2} \right], & r \leq r_0 , \\ \infty, & r > r_0 , \end{array} \right.
\end{equation}
where the spring constant is set to $k=30$ and the maximum spring length to $r_0 = 1.5$.
$NVE$ simulations were performed at the \emph{bead} number density $\rho_\mathrm{starpoly}=0.4$, and temperature $T_\mathrm{starpoly}=1$, under periodic boundary conditions for $N_\mathrm{starpoly}=10,\dots,2048$ star-polymers.
Hence, the side length of a cubic simulation box is set to $L = ( N_\mathrm{c}N_\mathrm{starpoly} / \rho_\mathrm{starpoly} )^{1/3}$.
As in the simple fluid case, $\mathcal{N}=16384$ samples were calculated using $\Delta t=0.002$ for $\mathcal{T}=10^5\Delta t = 200$.
However, since the complex geometry of a star-polymer may cause slower equilibration, a longer period of equilibration, $\mathcal{T}_\mathrm{equil}=5 \times 10^5 \Delta t=1000$, was employed.

The SACF and $\eta(t)$ were computed as follows.
For each sample trajectory, $p_{xy}$ was collected at every five time steps to calculate the SACF until $t=100$ using a standard time-averaging procedure~\cite{KimBorodinKarniadakis2015}.
Moreover, a numerical time-integration of the SACF was performed using the trapezoidal rule to obtain $\eta(t)$.
Then, the sample means and standard deviations of both SACF and $\eta(t)$ over $\mathcal{N}=16384$ samples were calculated. 
The standard error, $\sigma$, in a sample mean was estimated by $\sigma=\sigma_\mathrm{sample}/\sqrt{\mathcal{N}}$ where $\sigma_\mathrm{sample}$ denotes the standard deviation over the samples.  
Figures~\ref{fig:scf_vis-LJ} and \ref{fig:scf_vis-starpoly} show the time profiles of the SACF and $\eta(t)$ for the LJ fluid ($N_\mathrm{LJ}=2048$) and the star-polymer melt ($N_\mathrm{c}=21$ and $N_\mathrm{starpoly}=1024$), respectively.
For the LJ fluid, the VACF, $\langle v_x(0) v_x(t)\rangle$, and time-dependent self-diffusion coefficient, $D(t)$, were also computed from the same procedure.
The time profiles of these quantities are, respectively, shown in the insets of panels~(c) and (d) in Fig.~\ref{fig:err-LJ}.

\section{\label{sec_statisticalErrors}Statistical Uncertainty}

In Section~\ref{subsec_stat_uq}, we perform a statistical uncertainty analysis for the LJ fluid and star-polymer melt models and validate the statistical uncertainty quantification formulas given in Section~\ref{subsec_stat_uq_exp}.
In Section~\ref{subsec_gpa}, we further examine the Gaussianity of the shear-stress process.

\subsection{\label{subsec_stat_uq}Estimation of Statistical Uncertainties}

Here, we compare the actual statistical uncertainty level observed in MD simulations with that in the theoretical prediction based on the GPA.
We report the normalized standard error, $\widetilde{\sigma}$, defined in Eq.~\eqref{normSE}.
Theoretically predicted error levels are computed from Eqs.~\eqref{eq:err-corrFunc} and \eqref{eq:err-transCoef} using MD data of the corresponding autocorrelation function.
We also compare the long-time growth of the error levels in $\eta(t)$ and $D(t)$ with that of the asymptotic expression~\eqref{eq:asymptCoef}.

\begin{figure}
    \includegraphics[width=\textwidth]{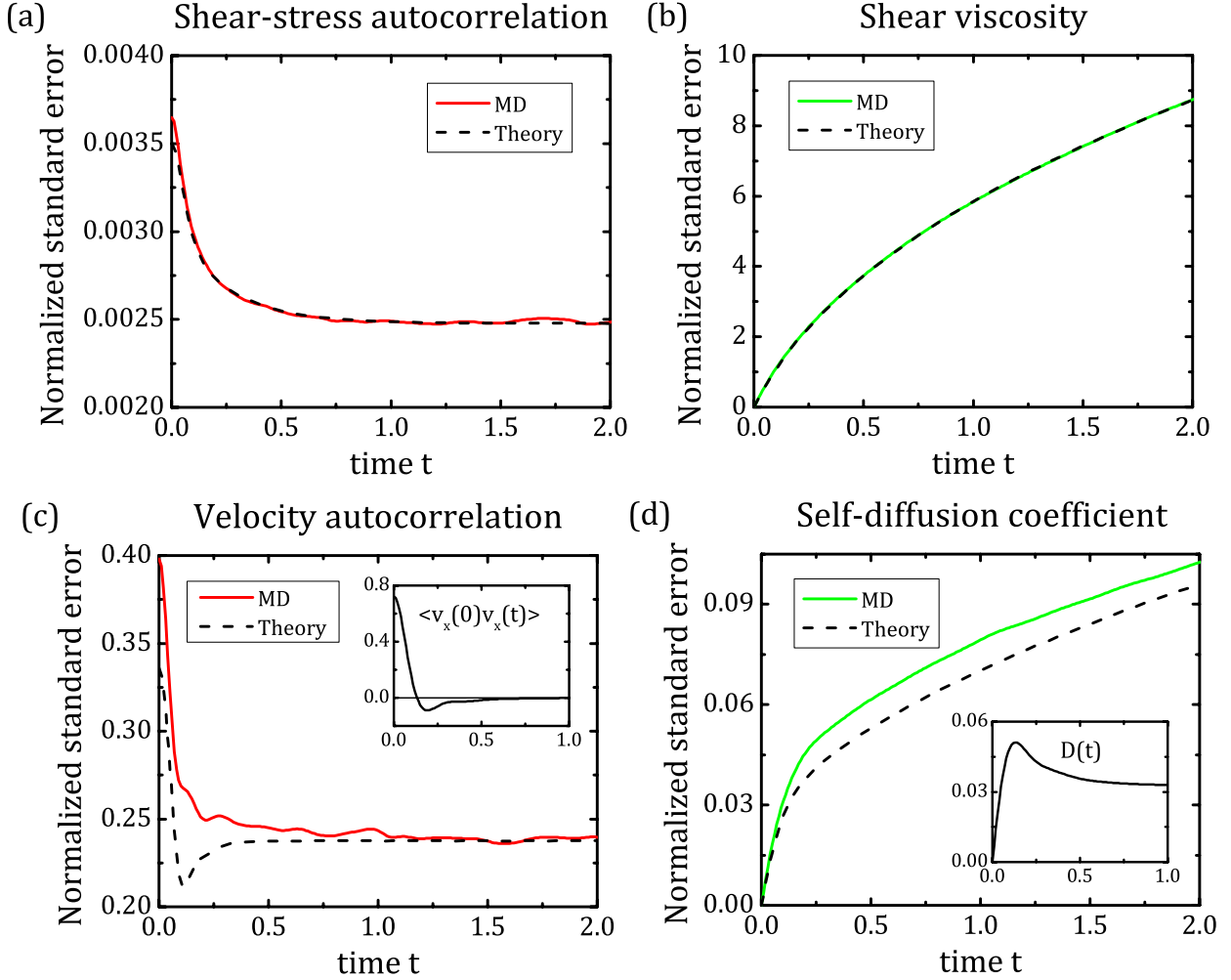}
    \caption{
    Comparison of the actual standard errors observed in MD simulations with that in the theoretical predictions for the LJ fluid ($N_\mathrm{LJ}=2048$).
    The normalized standard errors, $\tilde{\sigma} = \sqrt{\mathcal{NT}} \sigma$, are shown as the level of statistical uncertainty.
    The top row displays the shear viscosity cases (left: SACF; right: $\eta(t)$), whereas the bottom row shows the self-diffusion coefficient cases (left: VACF; right $D(t)$).
    The time profiles of the VACF and $D(t)$ are shown in the insets of panels (c) and (d), respectively.
    The SACF and $\eta(t)$ are shown in Fig.~\ref{fig:scf_vis-LJ}.
    }
\label{fig:err-LJ}
\end{figure}

Figures~\ref{fig:err-LJ}~(a) and (b) show statistical uncertainties involved in the shear viscosity estimation  of the LJ fluid with $N_\mathrm{LJ}=2048$.
For the SACF, the agreement between the MD result and theoretical prediction is remarkable.
The slight discrepancy at $t=0$ decreases with time and becomes negligible after $t=0.1$.
The standard error becomes constant around $t=1$ and its ratio to the initial error is approximately $1/\sqrt{2}$ as predicted by Eq.~\eqref{eq:asymptACF}.
For $\eta(t)$, the theoretical error estimate completely predicts the monotonic growth of actual error level with time.
As shown in Figs.~\ref{fig:err-LJ}~(c) and (d), theoretical error estimates for the evaluation of self-diffusion coefficient are fairly good but not as accurate as the shear viscosity case.
For the VACF, while the actual error level appears to monotonically decrease with time until attaining long-time values, the theoretical error estimate exhibits a dip at short periods due to the negative tail in the VACF (see the inset).
The latter underestimates the error level for $t<1$ but correctly predicts the long-time behavior.
For $D(t)$, the error level is underestimated, but its monotonic growth is predicted.

\begin{figure}
    \includegraphics[width=0.5\textwidth]{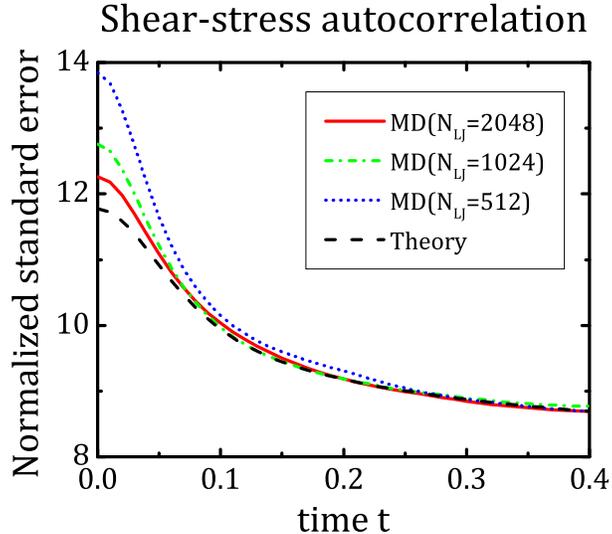}
    \caption{
    System size dependence on the statistical uncertainty level in the SACF of the LJ fluid.
    To compare MD results of three different system sizes, $N_\mathrm{LJ}=512$, 1024, and 2048, with the GPA-based theoretical predictions, the standard error is also normalized by the prefactor $ V (k_\mathrm{B}T)^{-1}$ (i.e., $\stackrel{\approx}{\sigma}=V(k_\mathrm{B}T)^{-1}\sqrt{\mathcal{NT}}\sigma$ is used).
    Since the theoretical predictions computed from the SACFs of the three system sizes are essentially the same, only the case with $N_\mathrm{LJ}=2048$ is shown.
    }
\label{fig:err-LJ_sizeEffect}
\end{figure}

Since the discrepancy between MD results and theoretical predictions is attributed to the violation of the GPA, the notable agreement observed in the shear viscosity case reveals that the Gaussianity is more pronounced in the shear-stress process.
A possible explanation is that the latter is represented by the collective dynamics of the whole system, see Eq.~\eqref{eq:virial}, whereas the velocity process only requires the information of a single particle.
This argument is based on a heuristic application of the central limit theorem.
One supporting observation is that theoretical error estimates for the shear viscosity become more accurate as the system size increases (see Fig.~\ref{fig:err-LJ_sizeEffect}), while those for the self-diffusion coefficient do not exhibit any system size-dependent improvement.
Hence, it is observed that the validity of the GPA for the shear-stress process is enhanced with increasing system size, which is consistent with our central limit theorem-based argument.

\begin{figure}
    \includegraphics[width=\textwidth]{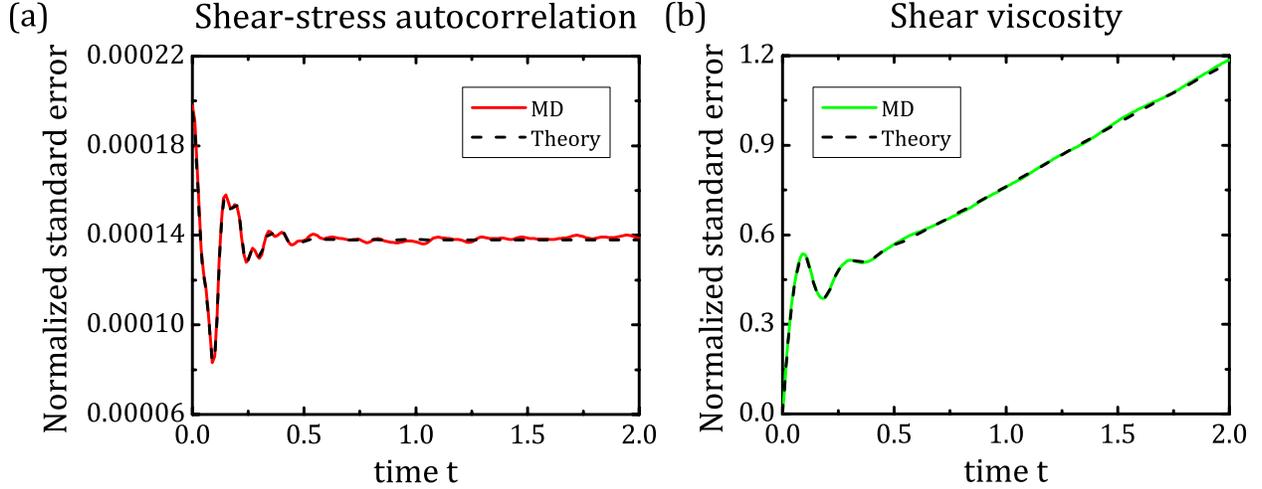}
    \caption{
    Comparison of the actual standard errors observed in MD simulations with the theoretical predictions for the shear viscosity estimation of the star-polymer melt ($N_\mathrm{c}=21$ and $N_\mathrm{starpoly}=1024$).
    The normalized standard errors of the SACF and $\eta(t)$ are shown in panels~(a) and (b), respectively.
    The time profiles of the SACF and $\eta(t)$ are shown in Fig.~\ref{fig:scf_vis-starpoly}.
    }
\label{fig:err-starpoly}
\end{figure}

Figure~\ref{fig:err-starpoly} shows that the structural complexity of a star-polymer molecule does not reduce the accuracy of the statistical uncertainty quantification formulas for the shear viscosity estimation.
However, the time profile of the error level itself becomes much more complicated due to the molecular structure.
That is, the complex short-time behavior of the SACF shown in Fig.~\ref{fig:scf_vis-starpoly}~(a) is reflected in the error level of the SACF and $\eta(t)$, and the growth of the error level in $\eta(t)$ is not monotonic at short periods.
In addition, as in the LJ fluid case, the enhanced accuracy of theoretical error estimates for a larger system is observed for all three sizes, $N_\mathrm{c}=11$, 21, and 31 of a star-polymer molecule.

\begin{figure}
    \includegraphics[width=0.5\textwidth]{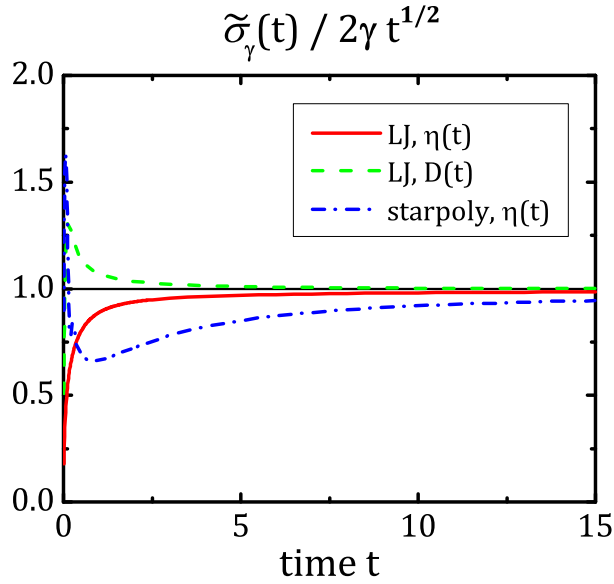}
    \caption{
    Long-time growth of the statistical uncertainty error level in the estimation of $\gamma(t)$.
    To check the validity of the asymptotic expression~\eqref{eq:asymptCoef}, the ratio, $\widetilde{\sigma}_\gamma(t)/(2\gamma\sqrt{t})$, is shown for the following three cases of $\gamma(t)$: $\eta(t)$ and $D(t)$ of the LJ fluid ($N_\mathrm{LJ}=2048$), and $\eta(t)$ of the star-polymer melt ($N_\mathrm{c}=21$ and $N_\mathrm{starpoly}=1024$).
    }
\label{fig:err_longtime}
\end{figure}

We finally observe the long-time growth of the standard errors in $\eta(t)$ and $D(t)$.
Since the growth is expected to be proportional to $\sqrt{t}$ under the GPA and more explicitly (see Section~\ref{subsec_stat_uq_exp} and the Appendix),   
\begin{equation}
    \label{eq:asymptCoef2}
    \sigma_\gamma (t) \approx 2\gamma\sqrt{\frac{t}{\mathcal{NT}}},
\end{equation}
where $\gamma$ is either $\eta$ or $D$, we show the time profiles of the ratio $\widetilde{\sigma}_\gamma(t)/(2\gamma\sqrt{t})$ in Fig.~\ref{fig:err_longtime}.
The ratio converges to unity in all three cases but the actual convergence time appears to depend on the time scale of $t^*$ satisfying $\gamma(t^*)\approx\gamma$. 
This is roughly estimated as 2 and 1 for $\eta$ and $D$ of the LJ fluid, respectively, and 10 for $\eta$ of the star-polymer melt.
We note that the ratio is already approximately unity at $t^*$ in all three cases, suggesting that the simple estimate~\eqref{eq:asymptCoef2} can be reasonable in a practical computation.

\subsection{\label{subsec_gpa}Verification of GPA}

We have so far shown the validity of the GPA from the consistency between MD results and the theoretical error estimates.
Here we examine it more directly.
That is, for the underlying processes $a(t)=p_{xy}(t)$ and $v_x(t)$, we perform two tests involving three- and four-time correlation functions.
These tests have been used in Refs.~\cite{KouXie2004} and \cite{KimBorodinKarniadakis2015}, respectively.

\begin{figure}
    \includegraphics[width=\textwidth]{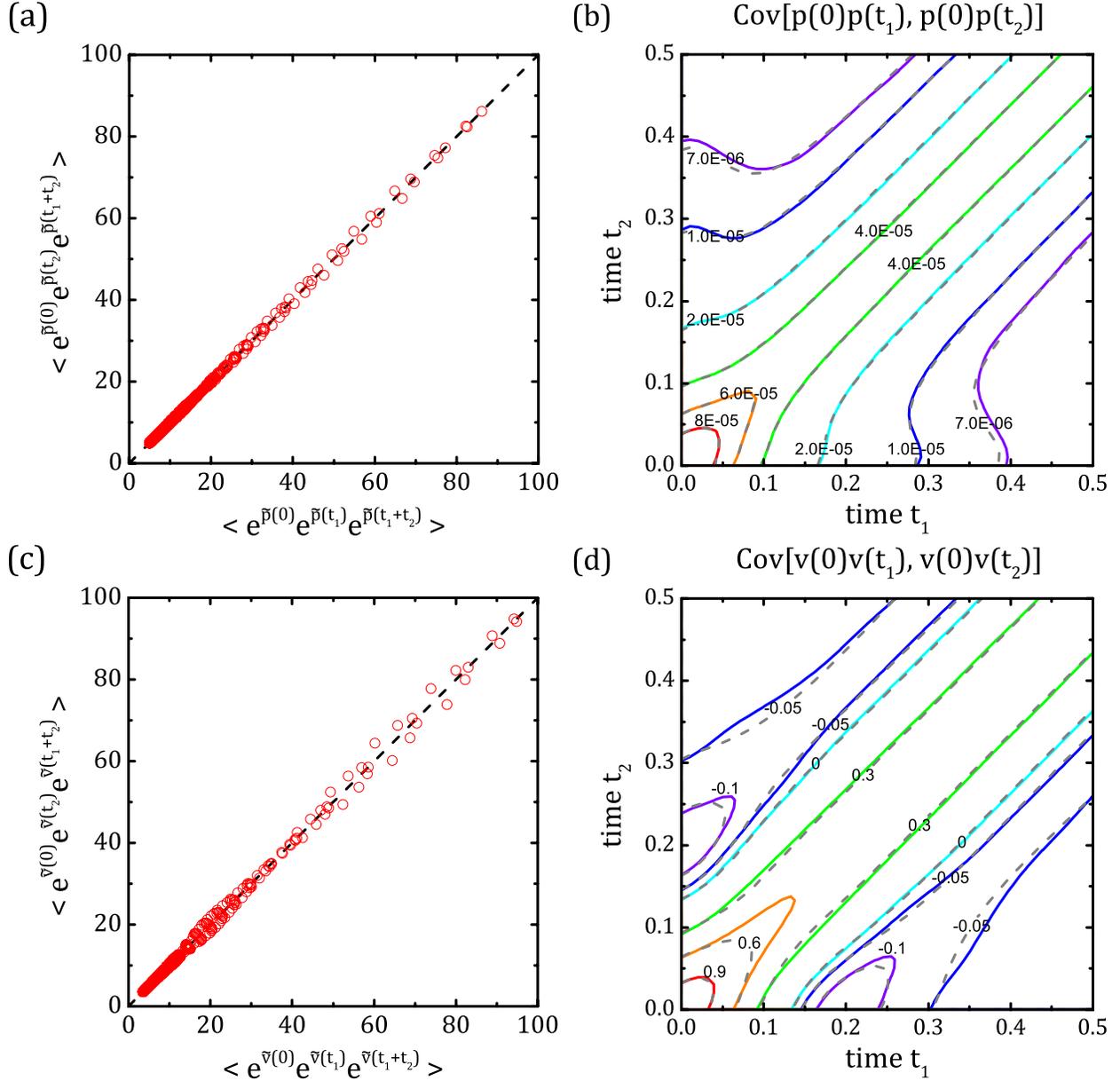}
    \caption{
    Verification of the GPA of $p_{xy}(t)$ and $v_x(t)$ using three- and four-time correlation function tests.
    In the left column, scatter plots of the three-time correlation functions in both sides of Eq.~\eqref{eq:3tCorr} are presented with the auxiliary lines $y=x$.
    In the right column, contour plots of the covariance in Eq.~\eqref{eq:cov-GPA} are shown.
    Colored solid lines depict the actual MD results, whereas dashed gray lines denote the GPA results.
    Numbers in the plots correspond to the values of each level curve.
    }
\label{fig:gpa}
\end{figure}

In the three-time correlation function test, the following property of a stationary Gaussian process $g(t)$ is used:
\begin{equation}
    \langle F(g(0)) F(g(t_1)) F(g(t_1 + t_2)) \rangle = \langle F(g(0)) F(g(t_2)) F(g(t_1 + t_2)) \rangle ,
\end{equation}
where $F(x)$ is an arbitrary function.
We choose $g(t)=\tilde{a}(t) \equiv a(t) / \sqrt{\langle a^2 (t) \rangle}$ and $F(x)=e^x$, and thus want to check whether
\begin{equation}
\label{eq:3tCorr}
    \langle e^{\tilde{a} (0)} e^{\tilde{a} (t_1)} e^{\tilde{a} (t_1 + t_2)} \rangle = \langle e^{\tilde{a} (0)} e^{\tilde{a} (t_2)} e^{\tilde{a} (t_1 + t_2)} \rangle .
\end{equation}
To this end, for the LJ fluid with $N_\mathrm{LJ}=2048$, three-time correlation functions in both sides of Eq.~\eqref{eq:3tCorr} are computed for various values of $t_1$ and $t_2$, and a scatter plot of the two correlation functions is drawn.
As shown in Fig.~\ref{fig:gpa}~(a) and (c), the two correlation functions lie near the line $y=x$, which proves the validity of the GPA of both $p_{xy}(t)$ and $v_x(t)$.
Also, better conformity is observed for the shear-stress process exhibiting more pronounced Gaussianity.

For the four-time correlation function, we choose the following covariance function and compare the MD results with the GPA results:  
\begin{equation}
\label{eq:cov-GPA}
    \mathrm{Cov} \lbrack a(0) a(t_{1}) , a(0) a(t_{2}) \rbrack \approx \langle a^2 (0) \rangle \langle a(0) a(t_{1} - t_{2}) \rangle + \langle a(0) a(t_{1}) \rangle \langle a(0) a(t_{2}) \rangle .
\end{equation}
In Figs.~\ref{fig:gpa}~(b) and (d), excellent agreement is observed for the shear-stress process, and overall good agreement is observed for the velocity process.
Hence, we reconfirm the Gaussianity approximation of the shear-stress process.

\section{\label{sec_finiteSystemSizeEffect}Finite System Size Effect}

In Section~\ref{subsec_sizeEffect}, we investigate the system size dependence of shear viscosity of both LJ fluid and star-polymer melt models using MD simulation results with well-controlled statistical uncertainty. 
Here we also discuss the system size effects on the self-diffusion coefficient of LJ fluid. 
In Section~\ref{subsec_kappaTilde}, we propose a physical entity that captures the length scale of the configurational rearrangement of the system and demonstrate its predictability for the estimation of system size effect on shear viscosity.

\subsection{\label{subsec_sizeEffect}Cooperatively Rearranging Regions}

\begin{figure}
    \includegraphics[width=\textwidth]{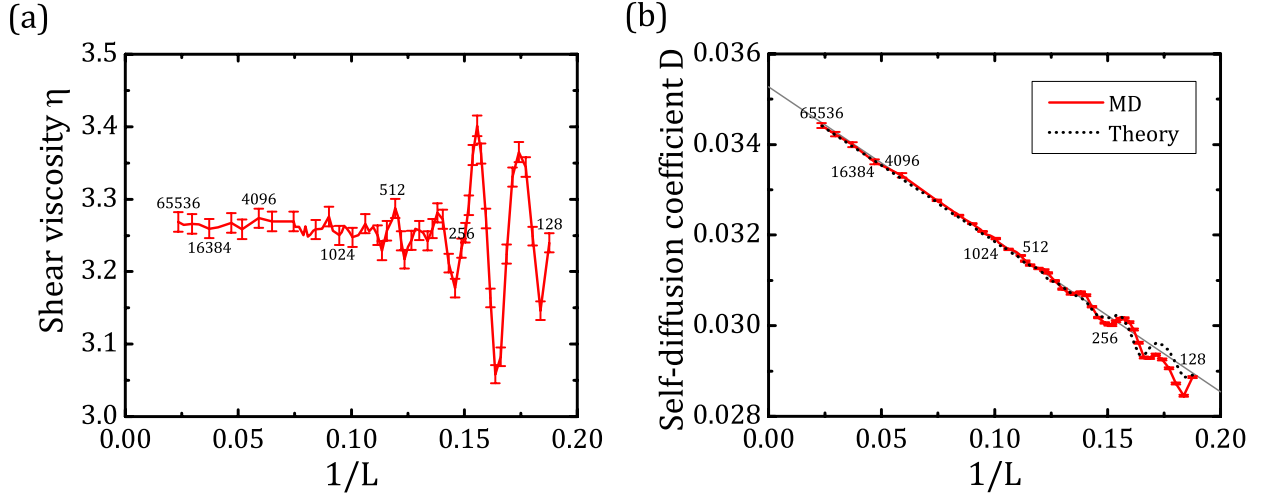}
    \caption{System size effects on (a) shear viscosity and (b) self-diffusion coefficient of the LJ fluid.
    Labelled numbers indicate the number of LJ particles in a simulation system.
    Error bars correspond to two standard deviations.
    The dotted line in panel~(b) depicts the theoretical prediction using Eq.~\eqref{eq:diffCoef} with the system size-dependent shear viscosities, whereas the grey solid line denotes the linear regression of the MD data.}
\label{fig:sizeEffect-LJ}
\end{figure}

Figures~\ref{fig:sizeEffect-LJ}~(a) and (b) present the system size effects on shear viscosity and self-diffusion coefficient of the LJ fluid.
No noticeable sign of scaling behavior is found for shear viscosity, whereas a clear $L^{-1}$ scaling behavior predicted by Eq.~\eqref{eq:diffCoef} is observed for the self-diffusion coefficient.
Shear viscosity is influenced by system size especially for small systems, resulting in a complex oscillatory behavior which dampens with increasing system size.
The oscillatory behavior becomes negligible above around $N_\mathrm{LJ}=1024$.
This behavior indicates the presence of a certain length scale of cooperatively rearranging regions, above which virtually all possible configurational rearrangements become allowed independently of the environment.
In other words, the complex oscillatory behavior of shear viscosity for small systems is a consequence of limited space for configurational rearrangements in dense fluids.
For self-diffusion coefficient, we observe an additional oscillatory behavior for small systems and identify this behavior to be associated with that of shear viscosity.
In fact, using the system size-dependent viscosities with Eq.~\eqref{eq:diffCoef}, we confirm that the size-dependent self-diffusion coefficients are accurately predicted for the entire range of the system size.

\begin{figure}
    \includegraphics[width=0.5\textwidth]{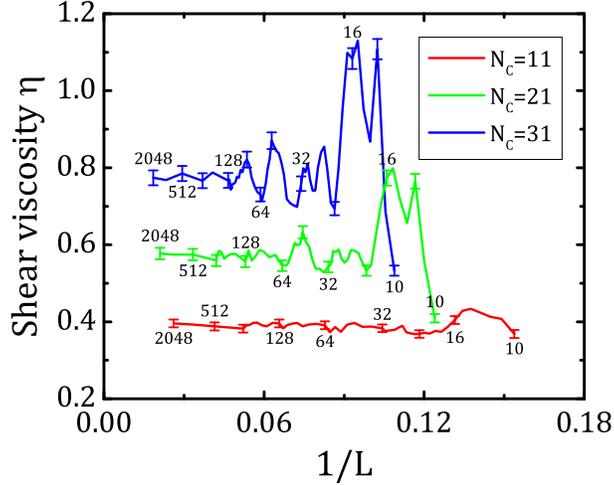}
    \caption{System size effects on shear viscosity of the star-polymer melt model.
    Three different sizes of star-polymers are compared: $N_\mathrm{c} = 11$, 21, 31.
    That is, star-polymers have $N_\mathrm{a} = 10$ arms with $N_\mathrm{b} = 1,2,3$ beads per arm, respectively.
    Labelled numbers indicate the number of star-polymer molecules in a simulation system.
    Error bars correspond to two standard deviations.}
\label{fig:sizeEffect-starpoly}
\end{figure}

Figure~\ref{fig:sizeEffect-starpoly} compares the system size effects on shear viscosity for the three star-polymer melt models with different values of arm length ($N_\mathrm{b} = 1$, 2, 3).
As in the LJ fluid model, an oscillatory behavior is observed for small systems.
Notably, the magnitude of oscillations and the length scale of cooperatively rearranging regions both depend on the effective size of star-polymers.
Since larger star-polymers can be considered as coarse-grained particles with larger effective radii, the length scale of cooperatively rearranging regions is expected to increase for larger star-polymers. 
Larger magnitude of oscillations and more complex patterns with increasing arm length are attributed to molecular interactions of complex molecular structures.

\subsection{\label{subsec_kappaTilde}Length Scale Estimation of Cooperatively Rearranging Regions}

We have so far observed that the system size dependence of shear viscosity, $\eta$, can be understood by the length scale of cooperatively rearranging regions, $\xi$.
Here we propose an entity, $\tilde{\kappa}$, from which one can accurately and easily estimate the length scale $\xi$.
It is defined as a normalized variance of shear-stress process,
\begin{equation}
\label{eq:normVar}
    \tilde{\kappa} = \frac{V}{k_\mathrm{B} T} \langle p_{xy}^2 \rangle.
\end{equation}
Note that $\tilde{\kappa}$ and $\eta$ are related to the SACF $C(t)$ as follows:
\begin{equation}
\tilde{\kappa} = \frac{V}{k_{\mathrm{B}}T}C(0),\quad \eta=\frac{V}{k_\mathrm{B}T}\int_0^\infty C(t) dt.
\end{equation}

\begin{figure}
    \includegraphics[width=0.5\textwidth]{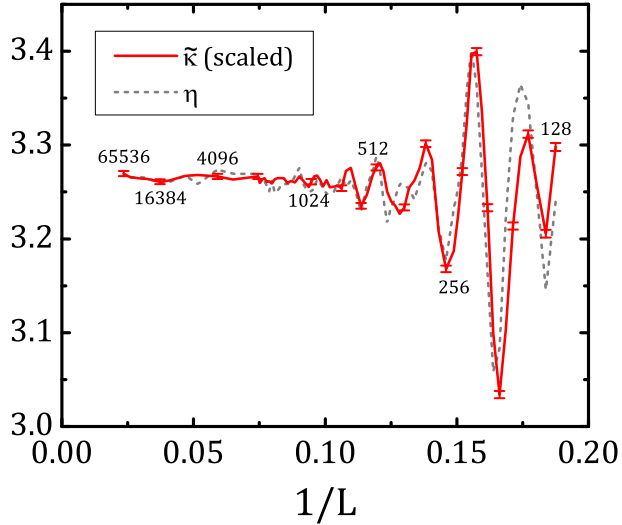}
    \caption{Comparison of the system size dependence of the normalized variance of shear-stress process, $\tilde{\kappa}$, and shear viscosity, $\eta$, of the LJ fluid.
    To visually compare the system size-dependent behaviors of the two quantities, we plot a scaled version of $\tilde{\kappa}$, $a \tilde{\kappa} + b$, using empirical parameters $a$ and $b$.
    The optimized values of $a$ and $b$ were obtained from the linear least squares regression.
    Labelled numbers indicate the number of LJ particles in a simulation system.
    Error bars correspond to two standard deviations.}
    \label{fig:kappaTilde}
\end{figure}

Figure~\ref{fig:kappaTilde} compares the system size dependence of $\tilde{\kappa}$ and $\eta$ for the LJ fluid.
As in $\eta$, the oscillatory behavior of $\tilde{\kappa}$ diminishes with increasing system size and becomes negligible when the system size is sufficiently larger than $\xi$.
The remarkable resemblance between the size-dependent behaviors of $\tilde{\kappa}$ and $\eta$ enables one to deduce some features of the system size effect on shear viscosity from those of $\tilde{\kappa}$.
This implies that the system size effect on shear viscosity of a dense fluid is largely determined by static equilibrium distribution and is mainly configurational.

In practice, $\tilde{\kappa}$ can minimize the efforts required to identify the system size effect on shear viscosity of a dense fluid.
That is, it is possible to compute an accurate value of $\eta$, without the need of any trial and error, by simply choosing a system size larger than the value of $\xi$ estimated from $\tilde{\kappa}$.
Since $\tilde{\kappa}$ is a \emph{static} equilibrium quantity, its evaluation is easier and faster compared to the computation of the dynamical quantity $\eta$. 
Moreover, its statistical error analysis is much simpler.
While Petravic's $\kappa$ (see Eq.~\eqref{eq:unrelaxedStress}) can be similarly used to estimate $\xi$, we note that periodic boundary conditions must be modified to induce unrelaxed stress~\cite{Petravic2004b}.
Contrary to the first-moment-based quantity $\kappa$, our second-moment-based quantity $\tilde{\kappa}$ can be defined and obtained under the standard periodic boundary conditions.

\section{\label{sec_conclusion}Conclusion}

We have investigated the statistical uncertainty and system size effect present in the equilibrium MD evaluation of shear viscosity, $\eta$, using the Green--Kubo formula.
Analytic expressions for statistical uncertainties in the SACF, $\langle p_{xy}(0)p_{xy}(t)\rangle$, and time-dependent shear viscosity, $\eta(t)$, were presented and verified through MD simulations of three-dimensional LJ fluid and star-polymer melt systems at relatively high densities.
It was observed that our statistical error estimates based on the GPA predict the actual error levels with remarkable accuracy.
By comparing with the case of self-diffusion coefficient, $D$, it was confirmed that shear-stress process, $p_{xy}(t)$, has more pronounced Gaussianity than the velocity process, $v_x(t)$, of a tagged fluid particle.
We explained this pronounced Gaussianity of $p_{xy}(t)$ by a central limit theorem argument; since the shear-stress process is expressed as a sum depending on the whole degrees of freedom (see Eq.~\eqref{eq:virial}), it is expected to be well approximated by a Gaussian process if there are sufficiently many degrees of freedom.  
This argument implies the enhancement of GPA for increasing system size, which was also confirmed in this work.

With controlled statistical uncertainties, we have verified that the shear viscosities of both systems exhibit strong size effects in small systems but without any overall scaling behavior.
The complex oscillatory behavior observed in the shear viscosity values of small systems was identified to result from limited space for configurational rearrangements.
Accordingly, the length scale of cooperatively rearranging regions, $\xi$, was estimated from the minimum system size where this behavior disappears and a reliable value of $\eta$ is obtained.
It was also observed that both the magnitude of the oscillation and length scale $\xi$ depend on the molecular structure of the fluid.
Besides the absence of scaling behavior in $\eta$ (as opposed to the $L^{-1}$ correction in $D$), these observations indicate that the main mechanism causing the size effects on $\eta$ of dense fluids has a configurational origin rather than a hydrodynamic one.
Hence, we proposed the normalized variance of the shear-stress process, $\tilde{\kappa}$, as a measure by which the minimum system size $\xi$ for a reliable value of $\eta$ can be readily estimated.

Further investigation for other fluid states is, however, required to gain a deeper understanding of the system size effect on the SACF and $\eta$.
It is well known that the potential contribution is dominant in dense fluid, whereas the kinetic contribution is influential in dilute fluid.
In this work, we mainly interpreted the size effects at relatively high densities based on structure relaxation.  
For a dilute fluid (e.g., rarefied gas), however, different system size effects are expected, which should be investigated based on hydrodynamic interpretation.  
For future work, we plan to investigate various fluid density regions, including dilute, intermediate, and dense fluids.
By decomposing the SACF into various components, we will investigate the hydrodynamic nature of $\eta$ and its configurational aspect to improve our understanding of viscous momentum relaxation in fluids.

\begin{acknowledgments}
This work was supported in part by the U.S.\ Department of Energy, Office of Science, Office of Advanced Scientific Computing Research, Applied Mathematics Program under Contract No.~DE-AC02-05CH11231, and Korea Advanced Institute of Science and Technology (KAIST), College of Natural Science, Research Enhancement Support Program under Grant No.~A0702001005.
It was also supported in part by the U.S.\ Army Research Laboratory and was accomplished under Cooperative Agreement No.\ W911NF-12-2-0023.
An award of computer time was provided by the ASCR Leadership Computing Challenge (ALCC) program.
This research used resources of the Argonne Leadership Computing Facility, which is a DOE Office of Science User Facility supported under Contract DE-AC02-06CH11357.
\end{acknowledgments}

\appendix*

\newcounter{eqn}
\setcounter{eqn}{\arabic{equation}}

\section{\label{appendix_asympt}Asymptotic Behavior of $\sigma_\gamma(t)$}

\setcounter{equation}{\arabic{eqn}}
\renewcommand{\theequation}{\arabic{equation}}

\begin{figure}
    \includegraphics[width=0.5\textwidth]{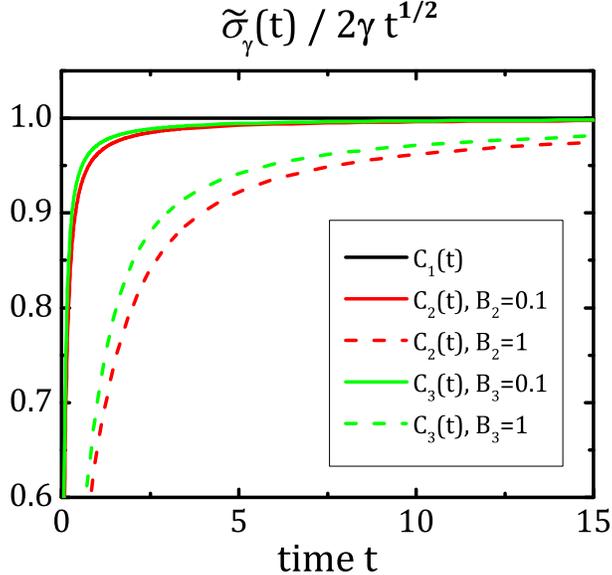}
    \caption{Time profiles of the ratio $\widetilde{\sigma}_\gamma(t)/(2\gamma\sqrt{t})$ for three model correlation functions, see Eq.~\eqref{eq:modelCt}.
    Results for two correlation times, $B_i=0.1$ and 1, are plotted for the exponential ($i=2$) and Gaussian ($i=3$) decays.    }
    \label{fig:append_asympt}
\end{figure}

We observe the long-time growth of the standard error $\sigma_\gamma(t)$ in Eq.~\eqref{eq:err-transCoef} and confirm the asymptotic expression~$\eqref{eq:asymptCoef}$.
To this end, we consider the following analytic forms of autocorrelation function $C(t)$, which are commonly adopted to model fluctuation correlations in molecular systems:
\begin{subequations}
\label{eq:modelCt}
\begin{align}
& C_1(t) = 2 B_1 \delta(t),\\
& C_2(t) = \frac{B_1}{B_2}\exp\left(-\frac{\lvert t\rvert}{B_2}\right),\\
& C_3(t) = \frac{2 B_1}{\sqrt{2\pi B_3^2}}\exp\left(-\frac{t^2}{2 B_3^2}\right).
\end{align}
\end{subequations}
Here $B_1$ denotes the time integral (i.e., $B_1 = \int_0^\infty C_i(t) dt$, $i=1,2,3$) and $B_2$ and $B_3$ correspond to the correlation times of $C_2(t)$ and $C_3(t)$, respectively (note that $B_1^{-1} \int_0^\infty t C_2(t) dt = B_2$ and $\left[ B_1^{-1} \int_0^\infty  t^2 C_3(t) dt \right]^{1/2} = B_3$).

For $C_1(t)$, the multiple integral in Eq.~\eqref{eq:err-transCoef} reduces to $4B_1^2 t$, yielding the normalized standard error $\widetilde{\sigma}_\gamma (t) = 2AB_1\sqrt{t}=2\gamma\sqrt{t}$.
Hence, we have $\widetilde{\sigma}_\gamma(t) / (2\gamma\sqrt{t}) = 1$ for all $t>0$.
For the exponential decay $C_2(t)$, we obtain
\begin{equation}
\widetilde{\sigma}_\gamma(t) = \gamma\sqrt{4t-3B_2+e^{-2t/B_2}(2t+3B_2)}
\end{equation}
and hence confirm Eq.~\eqref{eq:asymptCoef}.
For $C_3(t)$, it can be also shown that Eq.~\eqref{eq:asymptCoef} holds.
In the latter two cases with nonzero correlation times, it takes time for the ratio $\widetilde{\sigma}_\gamma(t)/(2\gamma\sqrt{t})$ to converge to unity.
As shown in Fig.~\ref{fig:append_asympt}, it takes more time for larger correlation time.

\bibliography{manuscript.bbl}

\end{document}